# Layer-by-layer water filling in molecular-scale capillaries

Mingwei Chen[1,2], Jingshan Wang[3], Artem Mishchenko[1,2], Ivan Timokhin[2], Fengchao Wang[3], Andre K. Geim[1,2], Qian Yang[1,2]

[1]National Graphene Institute, University of Manchester, Manchester, UK
[2]Department of Physics and Astronomy, University of Manchester, Manchester, UK
[3]Department of Modern Mechanics, University of Science and Technology of China, Hefei, China

Under ambient humidity, water spontaneously condenses in pores only a few nanometers in size, making nanoscale capillarity central to numerous natural phenomena and technological applications. At these dimensions, water may no longer be treated as a continuous fluid, yet the consequences of molecular discreteness for capillary condensation and filling remain poorly understood. Here we study nanocapillaries fabricated by van der Waals assembly and, using atomic force microscopy, monitor their wall deformations during humidity-driven water uptake. We observe two distinct regimes: layer-by-layer filling of flexible capillaries and abrupt filling of rigid ones. Flexible walls deform in steps of ~3 Å, corresponding to the sequential entry of individual water molecular layers. The different filling regimes are explained by the competition between deformation energy and oscillatory wall-water interactions. Our findings show that the molecular discreteness of water can profoundly affect ubiquitous capillary phenomena, with wall compliance selecting between discrete and abrupt filling.

**Introduction**

Capillary condensation—the spontaneous formation of liquid from vapor in confined geometries—underpins a wide range of processes under ambient humidity including friction, stiction, adhesion and lubrication[1-4]. In practice, at relative humidities (RH) commonly encountered in everyday environments, water condenses in pores, cracks, and contact gaps smaller than a few nanometers in size[3,5-8]. Because this spatial scale approaches molecular dimensions, the continuum description becomes questionable, as molecular-scale discreteness can strongly influence how condensation initiates and how the confined liquid advances during filling[3,5,6,9-12]. At the heart of the physics here is the known tendency of water to form stratified (layered) structures near solid interfaces and in thin films[13-16]. This interfacial stratification gives rise to oscillatory solvation (structural) forces[13,14,17-20] and can lead, for example, to discrete thickness plateaus and stepwise thinning[17,21]. In confined geometries, stratification of water (and other soft-matter liquids) can be driven by commensurability between the confinement size and characteristic molecular or aggregate length scales[13,14,18-20]. These known stratification effects raise a natural question for capillary condensation: when cavities and pores approach the molecular limit, does condensation-driven filling proceed through discrete states and, if so, under what conditions can such discreteness occur[3,5,10-12]? Addressing this question experimentally has been challenging because it requires both capillaries with precisely defined ~1-nm confinement and a quantitative, in situ readout of the condensation/filling process with angstrom precision.

This challenge has recently begun to be addressed using individual nanotubes[6,7,9] and two-dimensional (2D) capillaries[22,23]. Beyond revealing giant slip lengths[6,7,9], studies of water in narrow carbon nanotubes have uncovered pronounced anomalies attributed to packing of water molecules into quasi-1D structures[6,7,9,24-27]. Other work has reported dramatic changes in the dielectric properties of quasi-2D water[28], linked to violations of the ice rules that govern bulk behavior. Surprisingly, some experiments have still found quantitative agreement with classical continuum descriptions even under few-nm confinement. For example, the Washburn equation continues to describe capillary dynamics accurately in channels as narrow as 5 nm[29], while the Kelvin equation for water condensation has been shown to hold even for 1-nm slits[8]. In the latter case, the apparent agreement with classical behavior was attributed to the finite elasticity of the confining walls, which suppressed commensurability effects between the confinement size and the molecular diameter of water (~3 Å)[10,11]. This hinted that wall flexibility might be an important factor in whether molecular discreteness becomes apparent in capillary condensation.

In this report, we directly test this idea using 2D capillaries with different wall rigidity and find that wall compliance qualitatively changes the condensation/filling pathway. For sufficiently flexible walls, we observe 3-Å steps in capillary filling, consistent with the sequential entry of individual molecular layers of water. Our findings demonstrate that molecular discreteness and wall mechanics jointly control capillary condensation under ambient conditions, with implications for a wide range of ubiquitous condensation-driven phenomena.

**Capillary devices and AFM measurements**

The studied nanocapillaries were fabricated using van der Waals (vdW) assembly, as reported previously[8] and described in 'Device fabrication' in Supplementary Information (SI). In brief, multilayer graphene strips (number of layers, $N$) were sandwiched between two atomically flat graphite crystals, forming nanocapillaries with nominal height $h_s = Na$ and width $w \approx 150$ nm, where $a \approx 3.4$ Å is the thickness of monolayer graphene (Fig. 1 and Fig. S1). The nanocapillaries were placed on top of silicon-nitride (SiN) membranes fabricated from Si-SiN wafers (hereafter the entire assembled structures are referred to as 'nanocapillary devices'). The devices were then mounted to separate two chambers with controlled RH, as shown in Figs. S1 and S2a.

To monitor water filling, the nanocapillary walls were made flexible to allow measurable (Å-scale) deformations upon water imbibition. To achieve this, the top graphite crystal was chosen to have a thickness $H \approx 20\text{-}40$ nm. When empty, the nanocapillaries exhibited sagging $\delta$ of the top wall, which was induced by its vdW attraction to the spacer walls[8] and reduced the height at the nanocapillary center to $h = h_s - \delta$ (top inset of Fig. 1c). The degree of sagging depended critically on the top wall's bending rigidity which scaled as $\propto H^3/w^4$ (SI). If the top crystal was too thin ($H \lesssim 15$ nm) or the nanocapillaries were too wide ($w \gtrsim 200$ nm), they normally collapsed and strong vdW forces permanently pinned the sagged top wall to the bottom graphite[8,23]. Conversely, if the top wall was thicker than ~40-60 nm, sagging became too small for the resulting changes in $h$ to be measured with sufficient accuracy in our experiments. To study molecular-scale deformations associated with the entry of the first few water layers, the nanocapillaries were designed using such $N$, $H$ and $w$ that, after sagging, the empty channel central height $h(0)$ was expected to be $\lesssim 10$ Å. Due to local variations in adhesion between the top wall and spacers, different central regions of the same nanocapillary along its length ($y$-axis) exhibited various degrees of sagging, ranging from minimal to partial sagging and complete collapse. For each nanocapillary device, we carried out a painstaking search for regions where $h(0)$ was only a few Å, allowing initial room for no more than one or two molecular layers of water. Several regions with such $h(0)$ were typically found for each device (Fig. S1b), becoming the focus of further measurements.

Water filling was controlled using a two-chamber setup (Fig. S2a). The upper chamber, which provided AFM access to the capillary top surface, was maintained at low humidity (RH $\leq$ 10%) using a desiccant, while the lower chamber provided variable RH (see 'AFM measurements of capillary filling' in SI). AFM imaging of the top capillary walls was performed in PeakForce mode to monitor topography changes with high vertical precision (down to ~0.2 Å after spatial averaging; see 'Image analysis' in SI). All areas with suitable $h(0)$ were imaged simultaneously as RH gradually increased from ~10 to 90% in small steps (typically 2%). Each RH level was maintained for ~2 hours to ensure equilibration before proceeding to the next step (Fig. S2b). Two practical constraints defined the above RH range. Below 10% RH, no measurable changes were observed, consistent with the presence of an adsorbed water layer that cannot be removed at room temperature even at 0% humidity[2,30]. Above 90% RH, changes in $h$ were very small[8,23] and, additionally, AFM imaging became unstable due to a water meniscus frequently forming between the AFM tip and the graphite surface. The stiction is attributed to a wetting layer that spreads micrometers away from capillary openings into the dry AFM chamber and reaches the tip (Figs. 1a & S1b). Furthermore, we focused on measurements with rising RH, as water was found to evaporate from the nanocapillaries extremely slowly[8], preventing equilibration even after tens of hours and making decreasing-RH experiments impractical.

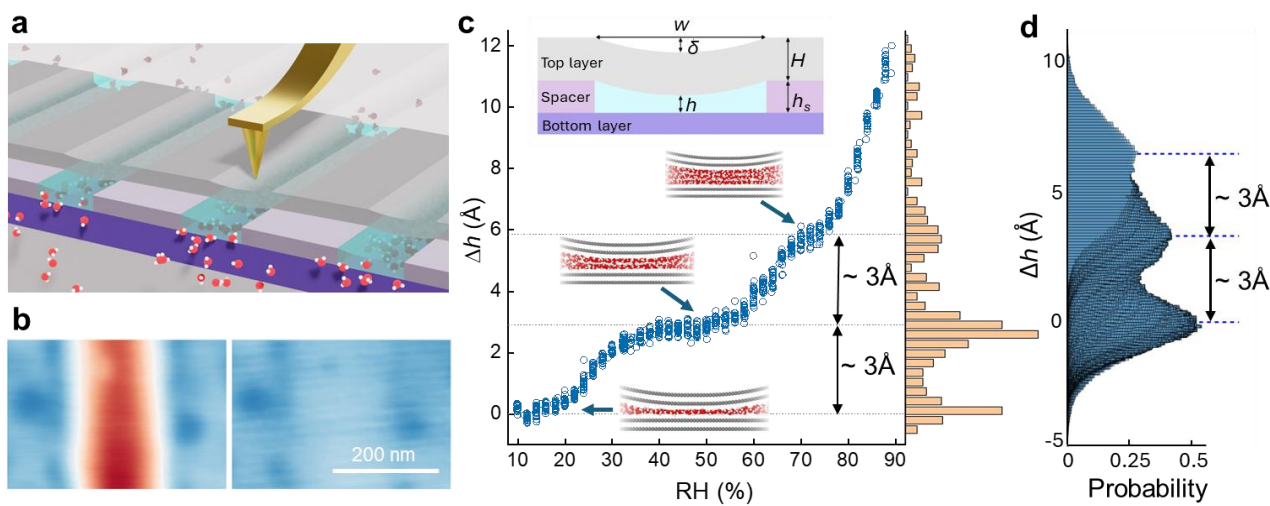

**Fig. 1| Layer-by-layer water imbibition**. **a**, Illustration of nanocapillary devices and their AFM probing. **b**, AFM images at low and high humidity (red-to-blue scale, 28 Å). At 10% RH (left), the top wall exhibits pronounced sagging, which practically disappears at 90% RH (right). **c**, Sagging $\Delta h$ as a function of increasing RH (blue circles) for a nanocapillary with $N = 6$, $H \approx 25$ nm, $w \approx 145$ nm. Three flattenings, spaced by ~3 Å (dotted lines), suggest sequential entry of water monolayers. The right panel shows a histogram of the $\Delta h$ values for the main-panel curve. Top inset: device schematic with annotated structural elements (the bottom graphite layer placed onto 500-nm thick SiN is assumed rigid). Other insets: illustrations of graphene nanocapillaries containing 1, 2, and 3 molecular layers of water. **d**, Each single-peaked histogram in this plot represents a distribution of $\Delta h$ values measured along a ~500-nm-long segment in the center of a nanocapillary ($N = 7$, $H \approx 28$ nm, $w \approx 160$ nm). Hundreds of such histograms measured at different RH are overlaid on top of each other. The resulting three peaks reveal that the nanocapillary segment spent most of the measurement cycle at heights separated by ~3 Å.

Each AFM measurement cycle yielded ~1,000 images, capturing the evolution of the nanocapillaries' water filling over the entire RH range. To extract quantitative information from these extensive datasets, we followed the protocol described in 'Image analysis' in SI. First, if more than 10% of images were compromised by a contamination buildup or scanning instabilities, we considered those measurements void and discarded the entire dataset because even modest gaps in the RH sequence made water-filling dynamics unclear (see, e.g., Fig. S4 where the gaps amount to <10% but still notably obscure the filling behavior). If discarded, the measurement cycle was repeated or, more often, a new device had to be fabricated. For successful cycles, AFM images were aligned to correct for thermal and mechanical drift during multiday measurements ('Image analysis'). Time-lapse videos compiled from aligned-image sequences demonstrate the effectiveness of our alignment procedures (Supplementary videos 1 and 2).

**Nanocapillary filling**

Using the aligned AFM images, we extracted heights $h$ in the central regions of selected nanocapillaries (with $h(0) \approx$ a few Å) and calculated changes $\Delta h = h(t) - h(0)$ as a function of time $t$ and RH. Several examples of the dependences $\Delta h(t,\text{RH})$ are provided in Figs. 1,2,S3 and S4 for nanocapillaries with flexible walls ($H < 30$ nm). In essence, the nanocapillaries swelled with increasing RH such that the top wall gradually lifted up and the capillary height smoothly increased. Fig. 1c shows this process for one of our devices with $N = 6$, where each blue circle corresponds to a $\Delta h$ value extracted from an individual AFM image. To ensure sufficient measurement accuracy, these values were averaged over a $\sim 20 \times 20$ nm$^2$ central regions of AFM images ('Image analysis' in SI). About 20 images were taken at each RH, providing multiple measurements of $\Delta h$ per RH level. These appear in Fig. 1c as vertical clustering of the blue circles, with their scatter reflecting temporal fluctuations in the measured height. Rather than averaging over the fluctuations at each RH, we chose to retain them. This presentation, complemented by histograms of the probability of observing particular $\Delta h$ over the entire RH sweep (right panel of Fig. 1c), better conveys both the experimental accuracy and details of the filling process.

Although nanocapillary swelling proceeded smoothly, pronounced flattenings appeared over certain RH intervals. Three plateau-like features are evident in Fig. 1c with the most prominent one spanning a wide range from 35% to 55% RH. The plateaus are separated by $\sim 3.0 \pm 0.2$ Å, consistent with both the intermolecular spacing in bulk water (3.1 Å) and the spacing typical for the first 1-3 water layers near graphite surfaces ($\sim 3$ Å)[11,13,15,31]. Perfect quantization is not expected due to averaging effects and varying layer separation in the stratified water (SI). It is also worth noting that the initial height $h(0)$ of the particular nanocapillary in Fig. 1c was also $\sim 3$ Å, which could imply that a monolayer of water was already present inside prior AFM measurements. This agrees with the expectation that water condenses in few-Å capillaries at ≲10% RH[8] and with the known fact that surfaces are invariably covered with adsorbed water monolayers even at zero humidity[1,2,30].

We also obtained histograms $\Delta h(y)$ where $\Delta h$ values were collected along a central $\sim 20$-nm-wide strip extending typically 200-500 nm along the $y$-axis of nanocapillaries. Over such distances, the sagging $\delta$ often varied considerably but, despite these local variations, we found that layer-by-layer imbibition involved the entire segment. Indeed, in Fig. 1d, individual $\Delta h(y)$ histograms taken at different RH for $\sim 500$-nm-long segment of a nanocapillary with $N = 7$ are overlaid, revealing three peaks that mark the most favorable positions of the top wall during the measurement cycle. These peaks are separated by $\sim 3$ Å, corroborating layer-by-layer filling across the entire imaged segment. Supplementary video 3 shows in detail how the $\Delta h(y)$ histograms evolved over time. Similar plateau-like features, separated by $\sim 3$ Å or its multiples, were observed in other devices with top walls thinner than 30 nm (see Figs. S3,S4 and S6 and Supplementary video 4 for further examples).

The described behavior differs qualitatively from that found previously[8] for similarly fabricated nanocapillaries with similar widths but thicker top walls ($H \approx 50$-$70$ nm). In those devices, $\Delta h$ remained unchanged at low RH and then exhibited a sudden jump at the condensation transition, reaching $\sim 10$ Å. Further increase in RH led to a gradual rise in $\Delta h$, attributed to a decrease in negative capillary pressure[8]. By contrast, the nanocapillaries described above showed no abrupt transitions but only smooth, stepwise filling. To confirm that this difference originates from different top-wall rigidity, we fabricated additional devices with $H > 30$ nm (SI), which displayed abrupt condensation transitions (Fig. 2b and Fig. S5). The contrasting behavior for top walls thinner and thicker than 30 nm is compared in Figs. 2a and 2b. In some cases, we also observed mixed behavior: smooth changes in $\Delta h$ at low RH followed by an abrupt increase in $\Delta h$ at higher RH (Fig. S6). These results link the present findings to those reported previously for stiffer walls.

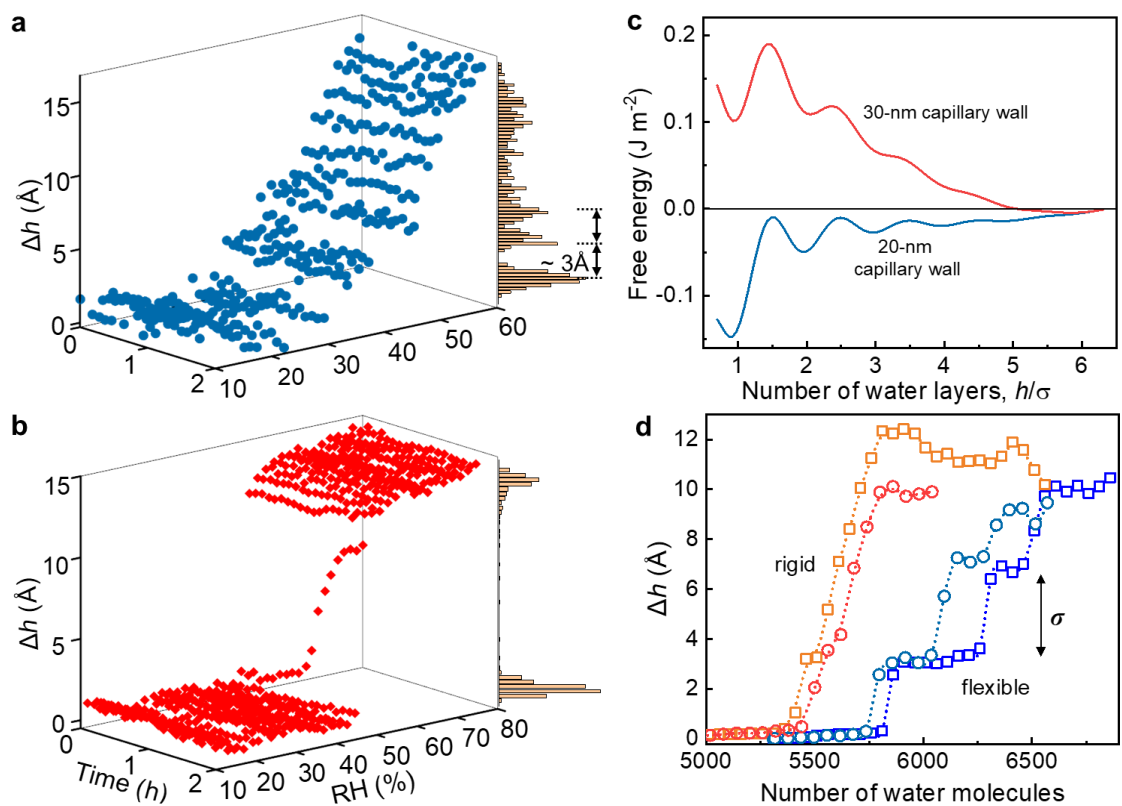

**Fig. 2| Gradual vs abrupt water filling. a-b**, Changes in the nanocapillary height shown as a function of both *t* and RH varied in 2% steps. The time axis emphasizes slow condensation dynamics at each RH step. Two types of condensation: (**a**) smooth filling with stepwise flattenings ($N = 6$, $H \approx 20$ nm, $w \approx 155$ nm) and (**b**) abrupt transition ($N = 8$, $H \approx 33$ nm, $w \approx 150$ nm). Each data point represents $\Delta h$ extracted from an individual AFM image. **c**, Calculated free energy of a model capillary with $N = 6$ as a function of number of water layers inside. Red curve: relatively stiff walls ($H = 30$ nm) where the elastic energy overcomes the oscillating solvation contribution. Blue curve: opposite case of flexible walls ($H = 20$ nm). Energy minima appear close to integer $h/\sigma$. All parameters used in the calculations are known either experimentally or from the literature (see text). **d**, MD simulations for flexible and rigid 2D capillaries. About 50 water molecules were added at each step, and the system was then allowed to relax for 0.1 ns (for details, see SI).

## Discussion

It is instructive to recall first how conventional nanoporous materials (with pores much larger than the molecular scale) respond to increasing RH. Water imbibition in such materials typically proceeds via two distinct stages[32,33]. At low RH, they undergo gradual expansion, known as the Bangham effect. This arises from adsorption of vapor molecules on internal surfaces, which reduces built-in surface stress, causing a net volume increase. At higher RH, once a critical threshold is reached, water spontaneously condenses to form menisci inside the nanopores, generating a large negative capillary pressure[1-3]: $P_c = -k_B T \rho_N \ln(\text{RH})$ where $k_B$ is the Boltzmann constant and $\rho_N \approx 3.3 \times 10^{28}$ m$^{-3}$ is the number density of water at room temperature $T$. For pores several nm in size, the resulting negative pressure can exceed 100 bar, and the meniscus-induced contraction often overwhelms the Bangham expansion. The latter is typically of the order of $10^{-4}$ in linear strain and, for 150-nm-wide capillaries, this would correspond to sub-Å deformations, below the resolution of our experiments. In sharp contrast, our molecular-scale capillaries exhibited continuous expansion – either stepwise or abrupt – without notable contraction, despite being subjected to much higher negative pressures (up to 3,000 bars at 10-20% RH)[2,8].

To rationalize the qualitatively different behavior, we consider a 2D capillary sketched in the inset of Fig. 1c. At the condensation transition, a meniscus forms and generates a capillary pressure $P_c = -2\gamma\cos\theta/h$ which pulls the top wall inward ($\gamma$ = 0.072 J m$^{-2}$ is the surface tension of water and $\theta$ is the contact angle). This pressure is opposed by the elastic restoring force $F = k(h_s - h)$ where $k$ is the effective spring constant of the top graphite wall. As shown in SI, $k$ can be approximated as $k = 512EH^3/125(1 - \nu^2)w^4$ with $E$ and $\nu$ being Young's modulus and Poisson's ratio of graphite, respectively. The two competing contributions to the free energy per unit area are then: elastic deformation energy $U_{el} = k(h_s - h)^2/2$ and capillary (meniscus) energy $U_{cap} = 2\gamma\cos\theta \ln(h/h_s)$. The latter expression assumes, for simplicity, a constant capillary height across the 2D capillary. The minimum in $U_{tot} = U_{el} + U_{cap}$ corresponds to the equilibrium state where the elastic force balances the capillary pressure and determines the top-wall position $h$ after the condensation transition occurs. This description represents the classical continuum framework for capillarity[1,2].

For capillaries that accommodate only a few molecular layers of water, the disjoining pressure energy[2] becomes significant, adding two extra terms to $U_{tot}$: entropic solvation energy $U_{ent}$, arising from the layered structure of water near surfaces, and vdW energy $U_{vdW} = - A/12\pi h^2$ which describes the attraction between opposing capillary walls ($A$ is the Hamaker constant of $\sim 1\times 10^{-19}$ J for graphite-water-graphite interfaces[2]). As a first approximation[2], the entropic term can be represented by an exponentially decaying oscillatory function $U_{ent} \approx -\Upsilon\cos(2\pi h/\sigma)\exp(-h/\sigma)$ where $\sigma$ is the thickness of a water monolayer and $h/\sigma$ therefore corresponds to the number of water monolayers inside the 2D capillary. To evaluate the coefficient $\Upsilon$, we carried out MD simulations (SI) which yielded $\Upsilon \approx 0.15$ J m$^{-2}$.

The competition between the four energy terms in $U_{tot}$ determines the water-filling behavior. For $h_s$ larger than several $\sigma$ (that is, for deep nanoscale but not molecular-scale confinement), both $U_{ent}$ and $U_{vdW}$ become negligible compared with $U_{cap}$, meaning that the standard capillarity behavior should recover[2,8]. In contrast, for molecular-scale capillaries, the disjoining energy can exceed $U_{cap}$. The entropic term is particularly important because it generates energy minima at integer $h/\sigma$, reflecting commensurability between the capillary height and the stratified water structure. Using $E \approx 1,000$ GPa, $\nu \approx 0.3$ and $\theta \approx 60°$ together with $w \approx 150$ nm and $H \approx 30$ nm, characteristic of graphite and our nanocapillaries, respectively, estimates show that $U_{el}$ becomes comparable to $U_{cap}$ when top-wall deformations reach $\sim$1 nm, which matches the scale of $\Delta h$ observed experimentally during water condensation. Using the same parameters, we also find that the nanocapillaries should collapse for $H \lesssim 15$ nm because vdW attraction between the top and bottom walls exceeds the restoring elastic forces ($U_{vdW} > U_{el}$ where $A \approx 6\times 10^{-19}$ J for graphite-gas-graphite interactions[2]), which matches well the thickness regime where we observed such collapse experimentally.

For further insight, Fig. 2c plots $U_{tot}$ for a nanocapillary with $N = 6$ and two representative $H = 20$ and 30 nm, corresponding to the nanocapillaries that exhibited stepwise and abrupt water filling, respectively. For the thicker wall, the strong $H^3$ dependence of $U_{el}$ pushes the $U_{tot}$ curve upward (red curve in Fig. 2c), leaving only a single stable state (energy minimum at $U_{tot} < 0$) at $h$ very close to $h_s$. This corresponds to the condensation transition accompanied by slight sagging, similar to the continuum description that neglects disjoining energy. For the more flexible wall ($H = 20$ nm), however, the combined capillary and disjoining pressures overwhelm the elastic restoring force, leading to negative $U_{tot}$ with pronounced energy minima, corresponding to 1, 2, 3 and 4 water monolayers inside the 2D capillary (blue curve). This implies layer-by-layer filling, which becomes continuous after 4 monolayers are present. The same curve also indicates that, at low RH, the capillary is likely to contain one monolayer of water (the deepest energy minimum). These predictions from the analytical model were found to be robust when varying $w$, $H$ and $h_s$ within the experimental parameter space as well as against uncertainties in the literature values of $A$, $E$, $\nu$ and $\theta$.

Although the model does not include the initial sagging caused by top and side wall interactions, nor does it account for variations in $h$ across the 2D capillary or possible parameter changes at the molecular scale (for example, $A$ probably changes significantly due to variations in the dielectric constant of nanoconfined water[28]), it nevertheless captures the essential physics. In particular, Fig. 2c reflects our key observations: (1) nanocapillaries with sufficiently flexible walls progress through a sequence of stable states corresponding to an integer number of water monolayers, and (2) even a modest (50%) change in wall thickness can determine whether condensation occurs stepwise or abruptly. In addition, if the initial height of a nanocapillary happens to fall between two energy minima (for example, $\sigma < h < 1.5\sigma$; see Fig. 2c) changes in RH may first drive the system into the low-$h$ equilibrium state. Consistent with this prediction, we occasionally observed a small (∼ 1 Å) but clear reduction in Δ$h$ during the initial RH increase (that is, a slight increase in sagging) which was followed by a 3-Å step to the next plateau (Fig. S3; Supplementary video 4). Taken together, the calculated energy landscapes reproduce remarkably well the key features observed experimentally.

To further support our analytical model, we simulated the filling process numerically (Fig. 2d; see 'MD simulations of water filling' in SI). A 2D capillary was modeled as two parallel plates (one bendable and one rigid, with fixed edges) capable of accommodating up to 4 water monolayers (Fig. S8). For sufficiently flexible walls, the capillary height evolved in quantized steps, reflecting an integer number of water monolayers in the central region of the 2D capillary (blue curves in Fig. 2d). By contrast, stiffer walls produced abrupt condensation transitions (red curves). The MD simulations are consistent with our experimental observations and, importantly, show that a slightly curved profile of the flexible wall (Fig. S8), which permits different numbers of water layers across nanocapillary's cross-section, does not obscure the layer-by-layer filling. This can be understood as follows: if the central part of a nanocapillary accommodates, for example, a single monolayer, it dominates the energetics, with contributions from other regions being comparatively small. Averaging therefore smears the energy minimum at $h/\sigma$ = 1 in Fig. 2c but does not eliminate it completely, in agreement with the observed stepwise collective changes along extended nanocapillary segments (Fig. 1d).

**Conclusion**

Our study shows that, in capillary condensation phenomena ubiquitous under ambient conditions, the filling pathway can depart qualitatively from the classical, century-old picture: instead of a single abrupt condensation transition, capillary filling can proceed through discrete, layer-by-layer states. The key insight is that wall compliance determines whether condensation remains effectively abrupt (despite the stratification of water under molecular-scale confinement), as expected in a continuum framework, or proceeds via sequential entry of molecular layers. While the stratification underlying the latter scenario has been well established, our results show how it can come into play during capillary condensation and filling. More broadly, our findings connect the molecular-scale interfacial structure in liquids with mesoscale mechanics in a way that is directly relevant to numerous ambient processes involving nanoconfined water, including adhesion/stiction, friction/lubrication and capillary-driven cohesion in granular media (e.g., sandcastles). The mechanism revealed here should generally apply to soft nanoporous materials and may also help guide strategies for engineering responsive nanoscale systems.